\documentclass[12pt,preprint]{aastex}
\usepackage{emulateapj5}
\usepackage{apjfonts}

\slugcomment{\underline{Accepted for}: ApJ \hfill \underline{Version}: \today}

\shorttitle{X-rays from thermonuclear SNRs}
\shortauthors{Badenes, Bravo, Borkowski \& Dom\'\i nguez}

\begin{document}

\title{Thermal X-ray emission from shocked ejecta in Type Ia Supernova Remnants.
Prospects for explosion mechanism identification}

%
%
%

\author{Carles Badenes,\altaffilmark{1,2} Eduardo Bravo,\altaffilmark{1,2} 
Kazimierz J. Borkowski,\altaffilmark{3} and 
Inmaculada Dom\'\i nguez\altaffilmark{4}}
\altaffiltext{1}{Dept. F\'\i sica i Enginyeria Nuclear, Univ. Polit\`ecnica de
Catalunya, Diagonal 647, 08028 Barcelona, Spain; badenes@ieec.fcr.es  
eduardo.bravo@upc.es}
\altaffiltext{2}{Institut d'Estudis Espacials de Catalunya, Barcelona, Spain}
\altaffiltext{3}{Department of Physics, North Carolina State University, 
Raleigh, NC 27695; kborkow@unity.ncsu.edu}
\altaffiltext{4}{Depto. F\'\i sica de la Tierra y del Cosmos, Universidad de Granada,
Granada, Spain; inma@ugr.es}

\begin{abstract}
The explosion mechanism behind Type Ia supernovae is a matter of continuing
debate. The diverse attempts to identify or at least constrain the
physical processes involved in the explosion have been only partially
successful so far. In this paper we propose to use the thermal X-ray
emission from young supernova remnants originated in Type Ia events
to extract relevant information concerning the explosions themselves.
We have produced a grid of thermonuclear supernova models representative
of the paradigms currently under debate: pure deflagrations, delayed
detonations, pulsating delayed detonations and sub-Chandrasekhar explosions,
using their density and chemical composition profiles to simulate
the interaction with the surrounding ambient medium and the ensuing
plasma heating, non-equilibrium ionization and thermal X-ray emission
of the ejecta. Key observational parameters such as electron temperatures,
emission measures and ionization time scales are presented and discussed.
We find that not only is it possible to identify the explosion mechanism
from the spectra of young Type Ia Supernova Remnants, it is in fact
necessary to take the detailed ejecta structure into account if such
spectra are to be modeled in a self-consistent way. Neither element
line flux ratios nor element emission measures are good estimates
of the true ratios of ejected masses, with differences of as much
as two or three orders of magnitude for a given model. Comparison
with observations of the Tycho SNR suggests a delayed detonation as
the most probable explosion mechanism. Line strengths, line ratios,
and the centroid of the Fe K$\alpha$ line are reasonably well
reproduced by a model of this kind. 
\end{abstract}

\keywords{hydrodynamics --- nuclear reactions, nucleosynthesis, abundances --- 
supernovae: general --- ISM: individual (SN 1572) --- supernova remnants ---  
X-rays: ISM}

\section{Introduction}

Type Ia supernovae (SNIa) play an important role in the chemical evolution
of galaxies, originating most of the Fe group elements in the interstellar
medium, and they have become a key tool in our understanding of the
universe by providing evidence for its accelerated expansion \citep{p99}. Still,
our knowledge of the physical
processes involved in the actual explosions is far from being complete.
Fundamental issues such as the mass of the white dwarf (WD) at the
moment of the explosion, the location of the ignition and the propagation
mode of the burning front have not been established yet, and a number
of different models or paradigms are capable of reproducing with reasonable
accuracy the optical spectra and light curves of observed SNIa and
their fundamental physical properties \citep[for reviews]{bk,hn}.

Given this situation, it is important to explore all the potential
sources of information that can shed some light on the detailed workings
of thermonuclear supernovae. The thermal X-ray emission from the shocked
ejecta in young supernova remnants (SNRs) originated by these explosions
is, in principle, well suited for this purpose: as the reverse shock
advances into the ejecta, it compresses and heats them to X-ray emitting
temperatures, producing spectra that depend on the composition, density,
temperature and ionization state of the material. The calculation
of the expected thermal X-ray emission from the shocked ejecta synthesized
in the different explosion paradigms, however, is not straightforward.
The propagation of the shocks and the density of the shocked gas is
governed by the dynamic evolution of the SNR, which at early stages
depends strongly on the ejecta density profile and is not well represented
by similarity or unified solutions \citep{tm}. The densities
involved are so low that the ensuing shocks are collisionless, and
their physics, and particularly how the postshock internal energy
is distributed among ions and electrons, is not well understood. Moreover,
the ages of young SNRs (up to a few thousand years) are short compared
with the time scales for thermal equilibration between ions and electrons
and for the onset of collisional ionization equilibrium, and therefore
the plasma is in a transient ionizing state that is difficult to calculate.

Our goal is to provide observers with the means to make meaningful
analysis of the thermal X-ray spectra of young, ejecta-dominated type
Ia SNRs and eventually to constrain the nature of the event that originated
them. Thus, here we present the first detailed models for these remnants,
evolved from a set of theoretical calculations of thermonuclear supernova
explosions. We have focused our attention on the general properties
of each explosion model, and their consequences on the emitted X-ray
spectra. A detailed analysis of particular objects is a delicate task
that requires fitting many observables and an exhaustive exploration
of the model parameter space in each case. This task is left for forthcoming
publications. In section \ref{explosion models} we give the relevant
characteristics of the explosion models in our grid, which includes
all the paradigms currently under discussion. The SNR simulation scheme
is described in section \ref{scheme}: hydrodynamics, plasma ionization
and heating, and synthetic spectra. Section \ref{results} contains
a discussion of our results and the relevant observable quantities
that can be derived from the models, as well as a preliminary comparison
with observations for Tycho SNR. The conclusions of this study are
presented in section \ref{conclusions}.

\section{\label{explosion models}Supernova explosion models}

In order to be able to discriminate between different explosion mechanisms
it is important that all the models be calculated consistently, i.e.
with the same physics included in the same hydro and nucleosynthetic
codes. With this requirement in mind, we have computed a grid of thermonuclear
supernova explosion models which is representative of the whole diversity
of theoretical models currently under debate. All the calculations
have been performed in one dimension, assuming spherical symmetry.
For each model, we have followed the explosion with a supernova hydrocode
until $10^{6}$\,s after the ignition, when the expansion has reached
a nearly homologous ($r\propto v$ everywhere) stage. 
The deposition of the energy of radioactive 
decay of $^{56}$Ni on the ejecta is taken into account in this supernova code.
After $10^{6}$\,s, however, the $^{56}$Ni disintegration becomes dynamically 
irrelevant 
because: 1) most of it has already decayed to $^{56}$Co, and 2) an increasing
fraction of the energy of the photons escapes the supernova due to the drop in 
opacity caused by the expansion. 
In addition
to the detailed nucleosynthesis, we have computed the light curves
in order to be able to compare with historical supernovae. The codes
used in the explosion calculations are the same as in \citet{b96}. More details 
are given in the appendix.

The different categories of SNIa explosion models that we have included
in our grid are:

\begin{itemize}
\item DET: Pure detonation model. In this class of models, the flame starts
close to the center of the WD, and propagates supersonically nearly
through the whole white dwarf, incinerating most of it to Fe-group
elements.
\item SCH: Sub-Chandrasekhar mass model. A He detonation is started at the
edge of a He envelope, which feeds a converging shock wave into the
C-O core. Close to the center, the converging shock wave transforms
into a C-O detonation which propagates outwards and processes the
rest of the core. A sandwiched structure is produced, rich in Fe-group
elements both in the inner and in the outer parts of the ejecta (in
our model, below a Lagrangian mass of $\sim 0.4\, \mathrm{M}_{\sun }$
and above $\sim 0.8\, \mathrm{M}_{\sun }$, in what was the He envelope,
see Fig. \ref{fig_composition}), and rich in intermediate-mass elements
plus C-O in between. 
\item DEF: Pure deflagration models. In these models the deflagration propagates
at the laminar flame velocity (a small fraction of the sound velocity)
close to the center, until the Rayleigh-Taylor instability develops,
deforms the flame surface, and accelerates the combustion. The flame
velocity remains subsonic all the way, but when the material has reached
velocities that are of the same order as that of the flame the expansion
quenches the flame. A large mass of unburnt C-O is ejected in the
outer layers.%
\footnote{Other deflagration models can be found in the literature in which
the mass of unburnt C-O is small while the mass of intermediate-mass
elements is large. This is the case, in particular, of the popular
W7 model \citep{nty}.
}
\item DDT: Delayed detonation models. In these models the flame propagates
initially as a slow deflagration, but a transition to a detonation
is induced at a prescribed flame density during the expansion of the
white dwarf. As a result, the otherwise unburnt C-O is processed partly
into Fe-group and partly into intermediate-mass elements.
\item PDD: Pulsating delayed detonation models. They differ conceptually
from delayed detonation models in that the transition to detonation
is induced only after the white dwarf has pulsed. The pulsation is
due to the inefficient burning produced by a slow deflagration, which
is unable to rise the whole energy of the white dwarf (gravitational
+ internal + kinetic) above zero.
\end{itemize}
Our model grid incorporates examples of all of these paradigms, and
it aims to cover the whole parameter space. For this paper we have
selected eight models from this grid as a representative sample, including
extreme cases of DEF, DDT, and PDD. Their properties are given in
table \ref{tab_models1}. There, $E_{\mathrm k}$ is the kinetic energy
of the ejecta, ${\mathcal M}_{\mathrm max}$ and $\Delta {\mathcal M}_{15}$ are, 
respectively,
the bolometric magnitude of the supernova at light curve maximum and
the change in bolometric magnitude between maximum and 15 days later,
and the rest of the columns are clearly defined by the header. The
meaning of the parameters in the second column is given in the appendix. 

In Figure \ref{fig_composition} we show the chemical structure of
each model after the short lifetime radioactive isotopes have completed
their decays (all the Fe that appears on the plots, for instance,
was synthesized as $^{56}$Ni in the explosion). Their density
structures are shown in Figure \ref{fig_densitystart}. In the delayed
detonation models, the transition from deflagration to detonation
happened at a Lagrangian mass of $\sim 0.2\, \mathrm{M}_{\sun }$, where
its imprint on the density profile can be seen. The chemical structure
is dominated by Fe-group elements up to $\sim 1.0\, \mathrm{M}_{\sun }$
for DDTa, at which point the flame density was too low to incinerate
matter to nuclear statistical equilibrium (NSE), leaving a buffer
of intermediate-mass elements. For DDTe this buffer is much larger,
because of the lower densities achieved by the detonation. The outermost
zone dominated by C-O is also larger than in model DDTa. The chemical
structure of the pulsating delayed detonation model PDDa is very similar
to that of DDTa, with the transition to detonation happening at $\sim 0.3\, 
\mathrm{M}_{\sun }$.
The different hydrodynamical histories of both models, however, are
reflected in their density profiles (see Fig. \ref{fig_densitystart}).
At the time of the transition, the density of the external layers
of PDDa was on average about two orders of magnitude lower than in
DDTa. As a result, PDDa displays a tail of very low density but larger
radii than DDTa. In model PDDe the transition to detonation took place
at $\sim 0.3\, \mathrm{M}_{\sun }$. Its chemical profile is similar to
that of DDTe, while its density profile is similar to that of PDDa.
In the deflagration model DEFa, the flame quenched due to the expansion
of the WD at a lagrangian mass of $\sim 0.7\, \mathrm{M}_{\sun }$, after
which a narrow ($\sim 0.1\, \mathrm{M}_{\sun }$ wide) region rich in
intermediate-mass elements was formed. The density structure is very
different from the delayed detonation models, due to the sudden termination
of nuclear energy generation which results in the formation of a bump
of unburnt material just above the quenching flame front. In model
DEFf the results were similar to DEFa, but the flame was quenched
at $\sim 0.9\, \mathrm{M}_{\sun }$.

\section{\label{scheme}Simulation scheme}

\subsection{Overview}

The density profiles from the explosion models described in the previous
section are used as an input for a standard hydrodynamics code (see
section \ref{hydrocode} for details) that follows the interaction
of the ejecta with a uniform ambient medium (AM). Hydrodynamical simulations
give the evolution of the radius, velocity, density and internal energy
per unit mass for each fluid element as a function of time, i.e.,
its dynamic history. This dynamic history, together with the chemical
composition of the fluid element obtained from the explosion models
is then used to calculate the evolution of the plasma heating and
the nonequilibrium ionization (NEI) in a self-consistent way (section
\ref{ionization code}), resulting in electron temperatures and ion
fractions for all the ions of the relevant chemical elements at each
time. The density, electron temperature and ionization structure within
a certain region of the SNR at a certain time then lead to a synthetic
or predicted thermal X-ray spectrum, which can be readily calculated
with a spectral code (section \ref{synthetic spectra}) and convolved
with any particular instrumental response in order to compare the
models with observations. In section \ref{scheme discussion} we will
review the approximations made at each stage of our simulation method
and ascertain its validity for the study of young, ejecta-dominated
Type Ia SNRs.

\subsection{\label{hydrocode}Hydrodynamics}

We have built a 1D Lagrangian hydrodynamics code such as described
by \citet{tm}, with an ideal gas equation of state ($\gamma = 
5/3$),
nonlinear pseudoviscosity and no external energy sources or sinks,
so the SNR model remains adiabatic throughout its evolution. We have
followed the interaction of the ejecta from various explosion models
with a constant density ambient medium (AM), with 
$\rho_{\mathrm AM}=10^{-24},\, 5\times10^{-25}$
and $10^{-25}$\,g\,cm$^{-3}$ as a sample representative of
the interstellar medium (ISM) conditions in most of the Galaxy and
the Magellanic Clouds.

\subsection{\label{ionization code}Plasma model and ionization calculations}

The theory for modeling transient heavy element plasmas in young SNRs
was developed by \citet[henceforth HS84]{hs}. These
authors describe the plasma with three Maxwellian populations: one
for the ions (including neutrals), and two for the electrons, a 'cold'
component for the electrons produced in the ionization process and
a 'hot' component for the electrons already present in the preshock
gas which were heated at the collisionless shock front. The relative
importance of the hot electron population thus depends on the amount
of collisionless heating at the shock, on the preshock ionization
state of the ejecta and on the composition of each layer, and in general
it will decrease with time as more 'cold' electrons are produced by
postshock ionization, especially if collisionless electron heating
is not efficient. The efficiency of this heating remains a controversial
issue: while theoretical calculations of the effect of plasma instabilities
on the electron population suggest a high degree of collisionless
heating at the shock \citep{cp}, this is difficult
to reconcile with observations of high velocity shocks, and this prediction
is being revised (see Laming 2000, for a discussion of collisionless
electron heating). We allow only for a small amount of collisionless
heating at the reverse shock, so we have represented the electrons
by a single Maxwellian population $T_{\mathrm e}$. In this two-fluid
model, the postshock temperature ratio between ions and electrons,
$T_{\mathrm e,s}/T_{\mathrm i,s}$, is a measure of the efficiency of 
collisionless
heating, varying between 0 (no heating) and 1 (total temperature equilibration).
For the models presented in this paper, we have assumed $T_{\mathrm
e,s}/T_{\mathrm i,s} = 0 
$, but we will explore the effect of different assumptions in future work.

All the elements in the unshocked ejecta are assumed to be in the
singly ionized state. A low preshock ionization state agrees with
the observational evidence provided by \citet{w83} for SN 1006,
one of the prototype galactic Type Ia SNRs, but we have found that
varying the charge of the unshocked ions has little impact in the
postshock ionization history, at least for charge states below ten.
For a more in-depth discussion of preshock ionization, see HS84.

In the shocked plasma, ions and electrons interact through Coulomb
collisions which gradually equilibrate the Maxwellian populations.
At the same time, the ongoing ionization of the ejecta increases the
electron population and therefore modifies the collision rate. Let
$\rho $ and $\varepsilon $ be the time-dependent mass density
and specific internal energy per unit mass of a given fluid element
as calculated with the hydrodynamic code, and $n_{\mathrm e},\, n_{\mathrm i}$
the number densities of electrons and ions, respectively. At any given
time, the total internal energy is distributed among ions and electrons
so that $\varepsilon _{\mathrm i}+\varepsilon _{\mathrm e}=\varepsilon $, and
the respective temperatures are related to the internal energies per
unit mass in each population by 
$\varepsilon _{\mathrm e,i}=(3kT_{\mathrm e,i}n_{\mathrm e,i})/2\rho  
$.
We can represent the chemical composition and ionization state of
the fluid element with the normalized number abundances for each element
X $f_{X}=n_{X}/n_{\mathrm i}$ and each ion q $f_{X^{q}}=n_{X^{q}}/n_{X}$
with $q=0$ for neutral and $q=Z_{X}$ for bare ions, so that
$\sum _{X}f_{X}=1,\, \sum _{q}f_{X^{q}}=1$. The plasma then evolves
according to the following equations (adapted from HS84): 
\begin{equation}
-\frac{d\varepsilon _{\mathrm i}/\varepsilon }{dt}=
\frac{d\varepsilon _{\mathrm e}/\varepsilon }{dt}=
\frac{1}{\rho \varepsilon }
\frac{2^{5/2}\pi ^{1/2}e^{4}n_{\mathrm e}n_{\mathrm i}\overline
{Z}^{2}\ln \Lambda }{m_{\mathrm e}\overline{A}m_{\mathrm u}k^{1/2}}
\frac{(T_{\mathrm i}-T_{\mathrm e})}{\left(\frac
{T_{\mathrm i}}{\overline{A}m_{\mathrm u}}+
\frac{T_{\mathrm e}}{m_{\mathrm e}}\right)^{3/2}},
\end{equation}
\begin{equation}
\frac{df_{X^{q}}}{dt}=
\frac{\overline{Z}\rho }{\overline{A}m_{u}}
\left[ 
I_{X^{q-1}}f_{X^{q-1}}+R_{X^{q+1}}f_{X^{q+1}}-
\left(I_{X^{q}}+R_{X^{q}}\right)f_{X^{q}}
\right],
\end{equation}
where $\overline{A}=\sum _{X}f_{X}A_{X}$ is the average mass
number, $\overline{Z}=\sum _{X}f_{X}\sum _{q}qf_{X^{q}}$ the
average ion charge in the fluid element, $m_{\mathrm u}$ the atomic mass
unit, $\ln \Lambda $ the Coulomb logarithm, all fundamental constants
take their usual values and all time derivatives are Lagrangian. $I_{X^{q}}$
and $R_{X^{q}}$ represent the ionization and recombination rates
\emph{from} ion q of element X, respectively, and are taken from the
recent compilation by \citet{m98}. Both equations are
obviously coupled in transient plasmas dominated by heavy elements
(where $d\overline{Z}/dt \neq 0$), and they have to be
integrated simultaneously. At the same time, the processes are controlled
by the precalculated variation of the hydrodynamic quantities $\rho $
and $\varepsilon $. The resulting stiff differential equation
system is solved with an implicit scheme, resulting in a time series
of temperatures and ion fractions for each fluid element. The values
of $T_{\mathrm e}$ and $T_{\mathrm i}$ are recomputed from the updated values
of $\varepsilon _{\mathrm e,i}$ and $n_{\mathrm e,i}$ after each new iteration
of equations (1) and (2). All the chemical elements with $f_{X}\geq 10^{-3}$
have been included in the calculations.

\subsection{\label{synthetic spectra}Synthetic spectra}

To produce the synthetic spectra, we have used an updated version
of the Hamilton \& Sarazin (HS) code incorporated in the XSPEC software
package \citep{arn}, which is suitable for modeling plasmas that
depart significantly from the collisional ionization equilibrium (CIE).
Given the density, ionization state and temperature of a fluid element
at a certain time, the HS code produces a predicted spectrum that
can be convolved with the response matrix of any appropriate instrument.
It is important to mention that this code relies on modern atomic
calculations for the Fe L shell emission from Li-like to Ne-like Fe
ions \citep{log}. This code includes no
atomic data for Ar, and has other limitations which restrict its applicability,
but problems of this kind are common to most of the currently available
NEI codes. The limitations of the spectral code should be considered
when comparing the predicted spectra with observations, but the quality
of our calculated spectra is satisfactory for our present purposes.

We will only discuss the integrated ejecta spectra: detailed models
for the shocked AM spectra and spatially resolved spectroscopy of
the ejecta are beyond the scope of this paper, even though they can
certainly be addressed using the simulation scheme we have just described.
We have used the response matrix for the CCD MOS-1 EPIC camera onboard
the XMM-Newton observatory, which provides an excellent spectral resolution
and is a reasonable choice for observing extended objects like SNRs.
The Doppler broadening of lines due to thermal and bulk motions have
not been included because their effects are small at the spectral
resolution of XMM EPIC-MOS.

\subsection{\label{scheme discussion}Simulation scheme discussion}

It is important to emphasize that our simulations assume spherical
symmetry for both the SNIa ejecta and the interaction with the AM. Our
aim is to show that young SNRs can be a valuable tool for studying
SNIa explosions, and the assumption of one-dimensional dynamics
is a first step towards that goal. In fact, it can be a satisfactory
approach in many individual cases, especially since Type Ia SNRs tend
to be, on average, more symmetric than those originated by core collapse
supernovae. In contrast to core collapse SNRs, they seem to retain the original
ejecta stratification (compare the results of Decourchelle et. al.
2001 on Tycho and Lewis et. al. 2003 on N103B with those of Hughes
et. al. 2000 on Cas A). Whenever deviations from spherical symmetry
affect the shape or the spectrum of a given object, a comparison with
our models might be useful as a description of the overall ionization
state and some bulk properties of the remnant, but it should be done
with care. Moreover, the effect that the Raleigh-Taylor instabilities
acting on the contact discontinuity (CD) might have on the ionization
and heating of the outer ejecta layers is difficult to ascertain.
As explained in \citet{wc}, the averaged density profile
is smoothed by the instabilities, but the fingers of shocked ejecta
retain a higher density than the surrounding shocked AM, so the results
of our one dimensional simulations discussed in the following sections
might not deviate much from a more realistic case. 

Even if spherical symmetry is assumed, it is worthwhile to note that,
while SNIa progenitor systems are not supposed to substantially
modify the surrounding ISM, the constant density hypothesis does not
agree with current presupernova evolution models. These models imply
a substantial modification of the AM, with profound effects in the
dynamic evolution of the remnant, as discussed in \citet{bb01}.
This possibility has not been taken into account here, but
will be explored in future work. 

Another issue that is of special concern is the validity of the adiabatic
hypothesis in the hydrodynamic calculations: whereas radiative losses
are generally not important for solar composition plasmas within time
scales of a few thousand years, heavy element plasmas will radiate
at a much faster rate, and the losses might have noticeable effects
on the dynamics much earlier in the evolution of the SNR. The problem
is complicated by the fact that losses are composition and ionization
state dependent, so they can only be evaluated \emph{a posteriori},
after the ionization calculations described in section \ref{ionization 
code}
are completed, and it is not possible to include their effect in the
hydrodynamics with our simulation scheme due to the fact that both
calculations are decoupled. This \emph{a posteriori} monitoring of
the radiative losses in a heavy element NEI plasma can be done using
the atomic data from \citet{sm} as described by \citet{l01}.
For evaluation purposes, we define the time scale for the
onset of radiative losses $t_{\mathrm rad}$ , as the time when the calculated
\emph{a posteriori} losses exceed 10\% of the internal energy in a
number of layers that amount to 5\% of the total ejecta mass. This
time scale is below 5000 years only for the deflagration models: $
t_{\mathrm rad,DEFa}=3.03\times10^{10}\, s;\, t_{\mathrm rad,DEFf}=
2.37\times10^{10}\, s$
for an interaction with $\rho _{\mathrm AM}=10^{-24}$\,g\,cm$^{-3}$, with
$t_{\mathrm rad}$ increasing for lower AM densities. Because radiative
cooling is a runaway process, the validity of these models beyond
$t_{\mathrm rad}$ is difficult to determine, but it is worthwhile to notice
that the losses are usually confined to a very small volume and the
effect on the overall dynamics of the SNR should not be important \citep{hss}.

\section{\label{results}Results}

\subsection{\label{hydro results}Hydrodynamics }

The interaction of Type Ia ejecta with a constant density AM, based
on a grid of one dimensional thermonuclear supernova explosion models, was
first explored in \citet[hereafter DC98]{dc}. In
that paper, the detailed dynamics of the interaction of six ejecta
density profiles from thermonuclear supernova models was examined
and compared with three analytical density distributions: an exponential,
a power law of index $n=7$ with a constant density core, and
a constant density profile. Except for the exponential, these analytical
functions have been widely used before to approximate SNIa ejecta
density as a function of velocity. Approximate temperature profiles
were also calculated assuming solar abundances for the supernova ejecta.
DC98 conclude that in all cases the density of the shocked ejecta
increases from the reverse shock towards the contact discontinuity,
that this density rise is usually coupled with a drop in temperature
in the area close to the CD and that the sharp structures in the ejecta
profiles, especially in the He detonation models, give rise to secondary
waves propagating in the interaction region which could affect the
instantaneous X-ray emissivity of the remnants.

When analyzing the ejecta-AM interactions, it is important to note
that, once the ejected mass $M_{\mathrm ej}$, kinetic energy $E_{\mathrm k}$
and density profile of an explosion model are fixed, the interaction
with the AM follows a scaling law for the AM density $\rho _{\mathrm AM}$ 
\citep{gul}.
We have tested this scaling law by computing the interaction of the supernova
models with the three different AM densities mentioned in section 
\ref{hydrocode}. Our results follow the scaling law up to the precision of the
hydrodynamic calculations. 
Therefore, it is sufficient to present the hydrodynamic
calculations for each model with a certain value of $\rho _{\mathrm AM}$
and then use the characteristic magnitudes defined by DC98 (eqs. [3], [4], and
[5]) to scale the results to any other $\rho _{\mathrm AM}$ value that
might be of interest.

\begin{equation}
R'=\left(\frac{M_{\mathrm ej}}{\frac{4\pi }{3}\rho_{\mathrm AM}}\right)^{1/3},
\end{equation}
\begin{equation}
V'=\left( \frac{2E_{\mathrm k}}{M_{\mathrm ej}}\right) ^{1/2},
\end{equation}
\begin{equation}
t'=\frac{R'}{V'}=\frac{M_{\mathrm ej}^{5/6}}
{\left( \frac{4\pi }{3}\rho_{\mathrm AM}\right) 
^{1/3}\left(2E_{\mathrm k}\right)^{1/2}}.
\end{equation}

In Figure \ref{fig_shockevol} we show our results for the ejecta-AM
interaction with $\rho _{\mathrm AM}=10^{-24}$g\,cm$^{-3}$. 
We have plotted the time evolution of the forward and reverse shock radii
($r_{\mathrm fwd},\, r_{\mathrm rev}$), the velocity of the forward shock 
$u_{\mathrm fwd}$,
the velocity of the reverse shock in the rest frame of the expanding
ejecta $u_{\mathrm rev}=(r_{\mathrm rev}/t)-(dr_{\mathrm rev}/dt)$ and the 
expansion
parameters for both shocks, defined as $\eta _{\mathrm fwd,rev}=d\ln 
(r_{\mathrm fwd,rev})/d\ln (t)$.
The time axis spans between 50 and 5000 years after the explosion,
and the reverse shock parameters have been plotted only up to the
time of rebound. 
In fact, our simulations of the interaction of the ejecta with the AM start at
$10^7$~s after the explosion, but we present our results only from year
50 on because,generally speaking, younger remnants are not expected to emit 
appreciably in X-rays.   

The dynamics of the forward shock is affected by
the differences in the density profiles during the first thousand
years, then all the models converge towards the Sedov-Taylor solution
($\eta _{\mathrm fwd}=0.4$); this transition leads to the change of slope
of the forward shock radii in the log-log plot (Fig. \ref{fig_shockevol}a).
The shock trajectories of the deflagration models (DEFa, DEFf) lag
behind the others because their $E_{\mathrm k}$ is lower, and they can
also be easily distinguished by their high $\eta _{\mathrm fwd}$ values
at early times, (Fig. \ref{fig_shockevol}e), about 50\% higher than
in the other models. Sudden increases in $\eta _{\mathrm fwd}$ can be
seen in the PDD models around $t=3\times10^{9}$\,s, and in the
SCH model at $t=6\times10^{8}$\,s (the rise is outside the range
of Figure \ref{fig_shockevol}e, but its effect can be seen at the
beginning of the plot) and $t=6\times10^{9}$\,s. In general, high
$\eta _{\mathrm fwd}$ values are found in models that have high density
material in the outermost ejecta, as the DEF models, or high density
layers in the outer ejecta preceded by a low density tail, as PDDa
at $\textrm{r}=1.6\times10^{16}$\,cm, PDDe at $\textrm{r}=1.4\times
10^{16}$\,cm,
and SCH at $r=1.9\times10^{16}$\,cm and $r=1.3\times10^{16}$\,cm
(see Fig. \ref{fig_densitystart}). These high density layers in the
outer ejecta transfer their momentum to the shocked material and to
the forward shock, leading to the increased $\eta _{\mathrm fwd}$ that
we have noted. The high density material, once shocked, stays close
to the contact discontinuity, and the low density tail, if present,
is also recompressed by the reflected shocks that ensue when the high
density layers are overcome by the reverse shock. The result is that
the density enhancement effect close to the contact discontinuity
is stronger for the DEF, PDD and SCH models than for the DDT and DET
models, where the momentum transfer is more gradual (for examples
of density maps see Fig. 3 in Badenes and Bravo 2003b). The dynamics
of some of these models are compared to those of the power law and
exponential analytical profiles in \citet{bb03a}. 

Summarizing, our conclusions agree almost completely with those of
DC98: the density of the ejecta always peaks towards the contact discontinuity
for all models at all times, and the rich internal structure of the explosion 
models produces a series of secondary shock waves that travel along the shocked
ejecta and AM reheating and recompressing the material. 
Our results also agree with DC98 in that the mean temperature profile drops 
towards the contact discontinuity, but we have found that this behavior of the
mean temperature cannot be extrapolated to the electron temperature, as is 
explained in the following section (see also Fig. 2 in Badenes and Bravo 2003b).
We have also found
that the models with dense layers in the outer ejecta, that is, the
deflagration models and, to a lesser extent, the PDD models and the
SCH model, tend to achieve higher densities in the region behind the
contact discontinuity. This will have a profound impact in the ionization
and temperature calculations.

\subsection{\label{emission measure}Temperature, ionization, and emission 
measure}

The spectral characterization of a young SNR is a complex issue: since
the density, ionization state, temperature and composition of each
fluid element are different, each region of the SNR will contribute
differently to the total integrated spectrum, and so will each chemical
element. A convenient way to measure these contributions is the emission
measure (EM) for element X, defined as
\begin{equation}
EM_{X}=\int _{V_{\mathrm sh}}n_{X}n_{\mathrm e}dV,
\end{equation}
where $V_{\mathrm sh}$ is the volume of shocked ejecta. For identical
physical conditions and a common history, elements with equal emission
measures contribute equally to the total integrated ejecta spectrum.
But the electron temperature $T_{\mathrm e}$ and ionization time scale
of the plasma, $\tau =\int n_{\mathrm e}dt$, which play a key role in
thermal NEI spectra, are different for each fluid element, resulting
in different spectra produced by fluid elements with identical emission
measures. \citet{blr} approached this problem for the
shocked AM in Sedov SNRs by introducing distribution functions, plots
of $T_{\mathrm e}$ and $\tau $ versus EM, but the use of this approach
for the shocked ejecta would call for an individual distribution function
for each chemical element due to the nonuniform chemical composition.
An incomplete, yet meaningful, description can be achieved by taking
the first moment of the distribution functions and calculate, for
each element X, an emission measure averaged electron temperature
$\left\langle T_{\mathrm e}\right\rangle _{X}$ and ionization timescale
$\left\langle \tau \right\rangle _{X}$. (Another quantity, an
ionization timescale averaged electron temperature is generally necessary
for a reliable modeling of X-ray spectra, but it is less important
than $\left\langle T_{\mathrm e}\right\rangle _{X}$ and $\left\langle \tau 
\right\rangle _{X}$
). We stress that we do not recommend single temperature, single ionization
timescale spectral models to fit shocked ejecta in young SNRs, we
are merely using the averaged quantities to describe average physical
conditions of various layers of ejecta in our models. In all the spectra
and calculations presented in the following sections we have used
the full profile of electron temperatures and ionization timescales
throughout the ejecta, not the averaged quantities.

The $EM_{X}$ evolution is shown in Figure \ref{fig_emevol} for
$\rho _{\mathrm AM}=10^{-24}$\,g\,cm$^{-3}$, between 20 and 5000 years after
the explosion. The contributions from different chemical elements
to the ejecta spectra depend strongly on both the composition profile
of the models and their dynamic evolution. The density enhancement
effect towards the contact discontinuity makes the chemical elements
in the outer layers of the ejecta more prominent than those in the
inner layers, so the importance of Fe in the model spectra is generally
much less than might be expected on the basis of a Type Ia elemental
composition alone. This relatively low prominence of Fe in spectra
of Type Ia SNR candidates might have been noticed on a number of occasions,
often accompanied by inordinately high apparent abundances of other
elements (see Hendrick, Borkowski, and Reynolds 2003 for the SNRs 0548-70.4
and 0534-69.9 and Lewis et al. 2003 and van der Heyden et al. 2002
for N103B). The estimated ejected masses of various elements, which
often indirectly rely on the assumption that chemical abundances are
proportional to the fitted $EM$ for each element in the spectrum,
are difficult to reconcile with the yields of theoretical explosion
simulations. This discrepancy between $EM$ and ejected mass is
most dramatic in the deflagration models (Fig. \ref{fig_emevol} e
and f), whose spectrum is completely dominated by C and O, with emission
measures of Fe about two orders of magnitude lower at all times, even
though the ejected mass of Fe is higher than that of C or O. Note,
however, that the peak value of the $EM_{\mathrm Fe}$ in the deflagration
models is about the same as in the other models, with the exception
of PDDa, and that radiative cooling could reduce considerably the
$EM$ of C and O in the deflagration models (see section \ref{scheme 
discussion}).
The prompt detonation model (Fig. \ref{fig_emevol}g) is the only
one whose spectrum is clearly dominated by Fe at all times, while
in DDTa and PDDa (Fig. \ref{fig_emevol}a and c) Fe takes over only
after a few hundred years, with important contributions of Si and
S (and O for DDTa) throughout the SNR evolution. The rise in the emission
measures around $t=10^{11}$\,s is due to the propagation of the
reverse shock after it rebounds at the center, reheating and recompressing
the ejecta; this rise happens earlier and is more gradual for the
elements in the inner layers than for those in the outer layers. Model
SCH is dominated by Fe only when the reverse shock is propagating
through the He detonation layer, afterwards O takes over, because
a second density enhancement region forms behind the interface between
this layer and the rest of the exploded WD.

Since the ionization and electron heating processes proceed faster
at higher densities, the corresponding 
$\left\langle T_{\mathrm e}\right\rangle _{X} $
and $\left\langle \tau \right\rangle _{X}$ plots in figures 
\ref{fig_teemevol}
and \ref{fig_tauemevol} are also affected by the enhancement towards
the contact discontinuity and the reverse shock rebound, albeit in
different ways. Those models with stronger density enhancement close
to the CD (DEFa, DEFf, PDDe, and, to a lesser extent, PDDa and SCH)
tend to have higher ionization timescales for most elements, and in
general, elements closer to the CD have higher $\left\langle \tau 
\right\rangle _{X}$.
In the deflagration models, for instance, C and O are always at a
higher $\tau $ and initially hotter than other elements. The
$\left\langle T_{\mathrm e}\right\rangle _{\mathrm C,O}$ drops at later times
because in these models the density is so high that electron-ion temperature
equilibration is achieved for most of the C and O in the ejecta before
$t=10^{10}$\,s and afterwards the electrons just cool due to adiabatic
expansion of this region of the SNR. Iron is generally hotter and
at a higher $\tau $ in the DDT and PDD models than in the DEF
models. The anomalous behavior of the plots for some elements (for
instance, Ni in DDTa and PDDa or Ca in SCH) is due to the averaging
in $EM$ and can be understood by comparing the curves with the
chemical composition profiles of Fig.\ref{fig_composition}. As the
reverse shock advances into regions with a much higher concentration
of a given element, the newly shocked (and therefore cooler and less
ionized) layers soon dominate the $EM_{X}$ and the averaged quantities
shift their values accordingly. It is worth noting that the enhanced
electron heating rate due to the higher densities towards the CD compensates
for the lower specific internal energies found in that region, so
we observe $T_{\mathrm e}$ profiles that always peak at the CD
(i.e. electrons and ions are closer to thermal equilibrium at the CD, whereas   
$T_{\mathrm e}\ll T_{\mathrm i}$ behind the reverse shock)
in contrast
to the behavior of the mean temperatures (DC98).

Our results for lower AM densities are similar, and we have found
that the hydrodynamical scaling laws (section \ref{hydro results})
also work reasonably  well for the emission measures, except at extremely
low AM densities. Scaled $EM_{X}$  agree to within a factor of
2, and so do the $\left\langle \tau \right\rangle _{X}$. The
temperature scaling seems more complex.  The deviations are accounted
for by the difference in average ionization state  which is expected
in SNRs that evolve in different $\rho _{\mathrm AM}$, because  hydrodynamical
scaling does not apply to ionization and electron  heating processes.
A detailed analysis of the ambient density effect on the spectral
characteristics of our models will be the subject of future work.

\subsection{Synthetic spectra}

The integrated synthetic spectra from the ejecta in our SNR models
are presented in figure \ref{fig_spectra} for 
$\rho _{\mathrm AM}=10^{-24}$\,g\,cm$^{-3}$,
500, 1000, 2000, and 5000 years after the explosion
(without interstellar absorption). As explained in the previous section,
the contributions from the different elements to the spectra depend
strongly on the details of the hydrodynamic evolution, and in particular
on the density enhancement towards the CD. Note how the spectra of
the deflagration models are always dominated by C and O, to the point
that their continua 'veil' the lines of the other elements at early
times. Even though their composition profiles are similar, PDDa and
PDDe have a richer line spectrum than DDTa and DDTe respectively,
because their mean ionization state is more advanced due to the higher
densities close to the CD (section \ref{hydro results}). The strength
of the Fe $K\alpha $ line varies from model to model, being important
at all times in DDTa, PDDa, DET and SCH, which have more Fe in the
outer layers of ejecta, while for DDTe and PDDe Fe K$\alpha $
is noticeable only after the first 1000 years, for DEFf after 2000
years and much later for DEFa. Oxygen can be easily identified in
the models with high $EM_{\mathrm O}$ and low $EM_{\mathrm Fe}$ (otherwise
its presence is partially masked by the Fe L complex): DDTe and PDDe
at early times, SCH up to 2000 years and the deflagration models at
all times. These oxygen bright models with weak Fe lines could be
easily mistaken for core-collapse SNRs. For the deflagration models
in particular, their strong C and O continua could make spectral analysis
difficult in presence of substantial interstellar absorption and in
view of calibration problems and low spectral resolution of CCD detectors
at low photon energies.

\subsection{Comparison with observations: the Tycho SNR}

A comparison with observations is essential for assessing validity
of theoretical models as a spectral characterization of young type
Ia SNRs and for learning about their progenitors. Our models are well
suited for qualitative comparison with observed spectra, but a quantitative,
detailed study of a given object is much more difficult. The excellent
quality of present day observations from XMM-Newton and Chandra has
surpassed our ability to model them accurately even with sophisticated
theoretical models, and additional issues such as modeling of shocked
AM spectrum, nonthermal spectrum subtraction, matching shock velocities,
expansion parameters and apparent radii, and spatially resolved spectroscopy
of the ejecta will have to be considered. Analysis of each particular
SNR might call for some modifications in the calculations or for a
more exhaustive exploration of the parameter space, and this task
is deferred to future work. However, a preliminary comparison of our
results with observations of a well studied SNR is very useful, as
it allows us to assess the strengths and limitations of our models.

The best object for this comparison is Tycho, the remnant of 
\objectname{SN 1572}.
Tycho is considered the remnant of a Type Ia supernova, and it has been
extensively studied in radio, optical, and X-rays. The X-ray emission
from the ejecta has been observed with high angular and spectral resolution
by XMM-Newton and Chandra \citep{d01,h02},
and spatially integrated but with moderate spectral resolution
by ASCA \citep[hereafter HG97]{hhp,hg}.
However, these extensive observational studies of Tycho resulted
in just a few comparisons with SNIa explosion models. \citet{imn},
and \citet{b89} calculated the predicted
X-ray spectra at the age of Tycho using the deflagration model W7
\citep{nty}, and compared it to the Tenma
and EXOSAT data to find that the observations could not be reproduced
by the W7 model unless substantial modifications to its structure
where introduced. 

We have made no attempt to reproduce the apparent size of the remnant
and the observed proper motion of Tycho; we just compared the X-ray
spectra of some of our models with the spatially integrated XMM spectrum
\citep[Figure 8]{d01}. The contribution
from the shocked AM has been modeled with the Sedov model in XSPEC
\citep{blr}, with mean shock temperatures and shocked
ISM emission measures derived from our hydrodynamical calculations
for each case (see Tab. \ref{tab_linestrengths}), and assuming no
collisionless electron heating at the shock ($kT_{\mathrm e}=0$), in
agreement with optical spectroscopy \citep{g01}. The
ionization age, defined as the product of postshock electron density
and the remnant's age, is $2.93\times10^{10}$\,s\,cm$^{-3}$. We
note that the shock speeds we obtain (and therefore the mean shock
temperatures $kT_{\mathrm sh}$ used in the Sedov models) are compatible
with X-ray measurements \citep{h00}, but not with optical or radio
observations \citep[and references therein]{h02}. The distance
to Tycho was assumed to be 2.3\,kpc, and we have used a $N_{\mathrm H}$
of $0.45\times10^{22}$\,cm$^{-2}$ as in HG97. 

A simple visual comparison of Tycho's spectrum with the model spectra
at t=500\,yr (Fig. \ref{fig_spectra}) reveals that one can eliminate
the DET model (absence of Si, S, and Ca lines) as well as DEFa, DEFf
and DDTe (mainly due to the absence of the FeK$\alpha $ line).
The contribution from the ISM (Sedov models) to the ejecta spectra
varies a lot depending on the type of explosion. It is quite large,
specially for the high energy continuum, in models DDTa, DDTe and
DET, and quite small in the others, except PDDa which is an intermediate
case. For the DEF models in particular, almost all the emission comes
from the ejecta, while for DET and DDTe the contribution to the underpredicted
lines from the ISM is small, and therefore these models remain unable
to account for the spectrum of Tycho. The spectra of the most promising
models have been plotted alongside the observed spectrum in Figure
\ref{fig_Tychospectra}.

For a more quantitative comparison we have computed the main characteristics
of the more prominent emission lines of our models, and compare them
to the observational results of HG97 in Tables \ref{tab_linestrengths}
and \ref{tab_lineratios}. The flux for the K$\alpha $ complexes
of Si, S, Ca and Fe was obtained by adding the contributions from
all lines due to atomic transitions from $n=2$ to $n=1$.

We concentrate now on models DDTa, PDDa, PDDe, and SCH, which have
not been eliminated previously. PDDa, PDDe, and SCH give too strong
Si and S Ly$\alpha $ lines, with respect to the corresponding
K$\alpha $ line complexes. These three models have in common
the formation of an extended high-velocity envelope (corresponding
to the accreted He envelope in the SCH model and to the external layers
detached from the white dwarf in the pulse previous to the formation
of the detonation in the PDDa and PDDe models), which is separated
from the rest of the ejecta by an abrupt change in density and followed
by dense layers rich in Si and S (see Fig. \ref{fig_composition}).
These dense layers remain dense when shocked, speeding up the ionization
processes inside them and producing more highly ionized Si and S and
therefore higher Ly$\alpha $/K$\alpha $ ratios. Other model
ratios compare reasonably well with the observed values, most of them
within a factor two to three. The Fe K$\alpha $ line centroids
show deviations larger than 50\,eV only for models PDDa and SCH. 

Given the limitations of the present work and the considerable uncertainties
in the atomic data it is impossible to reach a definitive conclusion,
but our preliminary results point to a delayed detonation explosion
as the progenitor of Tycho, probably similar to the DDTa model, with
a smooth density profile and a large amount of Fe in the outer ejecta.
Pure deflagration and pure detonation models can be confidently eliminated.
The only features that model DDTa does not reproduce well are the
Ca K$\alpha $ / Si K$\alpha $ ratio and the S K$\beta $
/ S K$\alpha $ ratio, suggesting that the ionization state of
S in Tycho may be a little bit higher than in the model. The angular
size of Tycho (8') implies a radius of 2.8($D$/2.3\,kpc)\,pc, while the
radius of the DDTa model at the age of Tycho is about 3.2 pc (see
Fig. \ref{fig_shockevol}). If the distance has been estimated correctly,
then $\rho _{\mathrm AM}$ must be higher than $10^{-24}$\,g\,cm$^{-3}$.
In simulations with a moderately higher value of $\rho _{\mathrm AM}$,
the ionization state of S in the DDTa model increases, bringing the
S K$\beta $ / S K$\alpha $ ratio closer to the observed
value while keeping the rest of the S and Si line ratios within reasonable
limits. A more detailed study of Tycho including these and other issues
will be the subject of future work.

\section{\label{conclusions}Conclusions}

We have performed self-consistent hydrodynamic and ionization state
simulations of the interaction of a grid of SNIa explosion models
with a uniform AM, producing a set of synthetic spectra for the integrated
thermal X-ray emission from the shocked ejecta in young SNRs. The
spectra show remarkable differences between the models: line strengths
and strength ratios, overall luminosity, shape of the continuum and
other parameters depend strongly on the type of supernova explosion that
gave birth to the SNR. These differences stem from the hydrodynamic
evolution, that is determined by the density profile of the ejecta
synthesized in the explosion, and from the evolution of the electron
heating and plasma ionization processes, which depend on both the
dynamic history and the chemical composition of the ejecta. A close
connection is thus established between supernova explosions and young SNRs,
implying that SNRs have the potential to become a new tool to discriminate
among various SNIa explosion models, a possibility that should
be tested by comparing models with observations. It is also clear
that any analysis of a thermal X-ray spectrum from the shocked ejecta
in a young SNR, whatever its type, has to make realistic assumptions
about the density and chemical composition profile for the ejecta,
because using uniform densities or homogeneous compositions is unrealistic
and will produce misleading results. The density enhancement effects
caused by the hydrodynamic evolution can lead to emission measure
ratios very different from elemental ratios (by number) in the material
ejected by the explosion. This is because the density enhancement
amplifies the contribution to the emitted spectrum by the elements
in the outermost layers of ejecta, where the differences between the
models are most pronounced. In contrast with the results of \citet{dc}
for the mean gas temperature, we find that both
density and electron temperature are highest at the contact discontinuity
in all cases.

We have made a preliminary comparison between our theoretical spectra
and the observed spectrum of the Tycho SNR. The best choice for Tycho
is a delayed detonation model with a large kinetic energy and a high
iron content. In particular, our delayed detonation model DDTa provides
reasonable agreement with observed line strengths, line ratios, and
Fe K$\alpha $ line centroid. Explosion models characterized by
high density layers in the outer ejecta preceded by an extended low-density
envelope, like pulsating delayed detonations and sub-Chandrasekhar
models give rise to an excess of ionization of Si and S, which leads
to high Ly$\alpha $/K$\alpha $ line ratios. Deflagration
models characterized by a large buffer of C and O in the external
layers would produce spectra with too weak lines and too strong continua.
One important result is that the emission measure ratios (as well
as the line ratios) cannot be safely used as a measure of the relative
abundances of various elements in the supernova ejecta; they can in fact
differ by as much as two or three orders of magnitude from the actual
abundance ratios.

A potentially important issue is related to various hydrodynamical
instabilities occurring during the explosion itself or shortly thereafter.
Multi-dimensional simulations of Type Ia explosions show the presence
of these instabilities \citep{kho,g03,n03,gb3,bg3}, which may lead to
inhomogeneities in freely
expanding ejecta. There is indeed observational evidence for the presence
of fast-moving ejecta clumps in Tycho SNR \citep{hg}.
But a quantitative understanding of the clumpiness of the freely expanding
ejecta is lacking at this time, both from theoretical and observational
perspective (see, for instance, Gamezo et al. 2003, and Thomas et al. 2002). 
A small degree of clumpiness is not likely to affect our
results based on 1-D simulations, but the situation will be different for
highly inhomogeneous ejecta. In extreme cases, dense clumps of
shocked ejecta might significantly contribute to spatially-integrated
X-ray spectra. The problem of clumpy supernova ejecta is clearly outside of
the scope of this work, and should be addressed in the future through 
multidimensional simulations and through detailed observational studies of
Type Ia SNR morphologies.

All the models, plots, and spectra shown here are available from the
authors upon request.

\acknowledgments

An important part of this research was conducted during a stay of
C.B. at North Carolina State University in the fall of 2002, funded
by GENCAT and NASA grant NAG 5-7153. We are grateful to Jordi Jos\'e
for his calculations and Pasquale Mazzotta for providing us with his
atomic data and routines. Una Hwang and Anne Decourchelle shared with
us helpful discussions. We are very grateful to Anne for providing
us with the XMM spectrum for Tycho, and to Martin Laming for his many
helpful suggestions about radiative cooling and other issues, and
his hospitality at NRL. C.B. and E.B. acknowledge funding from projects
AYA2000-1785 and AYA2002-04094-C03-01. K.J.B. acknowledges support
by NASA grant NAG 5-7153.

\appendix

\section{The Supernova Explosion Models}

In this appendix we give more details about the SNIa explosion models.
As has been mentioned before, the hydrocode, the nucleosynthesis,
and the light curve codes, as well as the physics included (equation
of state, nuclear reaction rates, etc) are the same as described by
\citet{b96}. The parameters of the models are given in Table
\ref{tab_models1}.

The detonation model was obtained from a $\sim 1.38\, \mathrm{M}_{\sun }$
WD in hydrostatic equilibrium (composed of equal masses of $^{12}$C
and $^{16}$O plus a 1\% by mass of $^{22}$Ne) whose internal
energy structure was adjusted to an adiabatic thermal gradient. The
ignition was initiated by incinerating the mass in the central layer,
and afterwards the detonation propagation was obtained consistently
by solving the hydrodynamic and nuclear evolutionary equations. Details
of a similar model can be found in \citet{b96}.

The sub-Chandrasekhar mass model was obtained from a WD formed by
a C-O core of $0.8\, \mathrm{M}_{\sun }$ surrounded by a He envelope
of $0.2\, \mathrm{M}_{\sun }$. This envelope was the result of He accretion
over the C-O core at a steady rate of 
$3.5\times10^{-8}\mathrm{M}_{\sun }$\,yr$^{-1}$. The hydrostatic evolution of 
the white dwarf subject to accretion
was computed by J. Jos\'e, who kindly provided us with the initial model
for the supernova explosion calculation (private communication). In this
initial model, the temperature at the base of the He envelope was
high enough to induce a spontaneous He detonation. The evolution past
this point was followed with the same hydrocode as above.

The rest of the explosion models started from the same initial configuration:
an isothermal white dwarf in hydrostatic equilibrium, with the same
chemical composition and central density as that used in DET model
(the differences in the thermal structures of both configurations
account for the small difference in total mass that can be seen in
Table \ref{tab_models1}). Explosion models starting from different
central densities $\rho _{c}$ (i.e. WD masses) do not produce
substantially different energies or light curves (with the exception
of a slight decrease of $^{56}$Ni yield with increasing central
density due to a larger electron capture rate, Bravo et. al. 1993),
so we did not consider variations in $\rho_{\mathrm c}$.

The flame propagation velocity in the pure deflagration models was
obtained as the maximum between the laminar flame velocity (as given
by Timmes and Woosley 1992 and updated by Bravo and Garc\'\i a-Senz 1999)
and the turbulent velocity, $v_{\mathrm RT}$. The turbulent velocity
was calculated as $v_{\mathrm RT}=\kappa r_{\mathrm fl}/\tau _{\mathrm RT}$, 
where $r_{\mathrm fl}$
is the flame radius, $\tau _{\mathrm RT}$ is the local Rayleigh-Taylor
time scale at the flame location, and $\kappa $ is a parameter
given in Table \ref{tab_models1} (see Bravo et. al. 1996 for details).
Furthermore, the mixing of matter and energy across the flame front
was limited as proposed by \citet{w87}. In this formulation,
the mixing is not allowed until the Rayleigh-Taylor front has propagated
to a prescribed fraction, $\theta $, of the mass of the shell
(in all our deflagration models we took $\theta =0.5$). Thus,
if the mass of the shell ahead of the flame front is $\Delta M$,
the mixing was started when the condition 
$\int ^{t}_{t_{0}}\frac{dM_{\mathrm RT}}{dt} dt=\theta \times\Delta M$
was held, with $t_{0}$ equal to the time of incineration of the
last incinerated shell, and with 
$\frac{dM_{\mathrm RT}}{dt}=4\pi r^{2}_{\mathrm fl}\times
\rho _{\mathrm fl}\times v_{\mathrm RT}$.
Once this condition was fulfilled, the transfer of internal energy
between the Rayleigh-Taylor unstable shells was allowed. Afterwards,
the flame propagation was obtained consistently by the consequent
increase in the nuclear energy generation rate and its feedback on
temperature. At densities below a few times $10^{7}$\,g\,cm$^{-3}$
the energy generated goes predominantly to create and maintain an
electron-positron pair gas rather than to increase the temperature,
which is the ultimate cause of the quenching of the flame. 

In the delayed detonation models the flame was propagated initially
as a deflagration. In that phase, the flame velocity was taken as
a constant fraction, $\iota $, of the local sound velocity (in
our models we took $\iota =0.03$). As the flame traveled through
lower and lower densities, the sound velocity and hence the deflagration
speed decreased, and the flame was eventually quenched. Afterwards,
the flame front was artificially accelerated up to a large fraction
of the sound speed. Following this fast propagation through a few
mass shells (typically 2-3 shells were enough) a detonation formed
and propagated through the rest of the star. The location of the transition
is univocally determined by the parameter $\rho _{\mathrm tr}$ (Tab.
\ref{tab_models1}), that is the density ahead of the flame at which
the sudden acceleration was imposed. In delayed detonation models
as well as in the pure detonation and sub-Chandrasekhar models, Rayleigh-Taylor
mixing was not allowed. The algorithm for the flame propagation in
pulsating delayed detonation models was the same as in delayed detonation
models, the only difference being that in PDD the detonation was not
triggered until the WD had pulsated. Further details (in particular
the nucleosynthetic output and more complete information about the
light curves) are available to interested readers upon request. 


\begin{deluxetable}{lcllllllllll}
\tabletypesize{\scriptsize}
\tablecaption{Properties of the explosion models\label{tab_models1}}
\tablehead{
\colhead{Model} &
\colhead{Parameter\tablenotemark{a}} &
\colhead{$M_{\rm ejecta}$} & 
\colhead{$E_{\mathrm k}$} & 
\colhead{${\mathcal M}_{\rm max}$} & 
\colhead{$\Delta {\mathcal M}_{15}$} &
\colhead{$M_{\mathrm C+O}$} &
\colhead{$M_{\mathrm Si}$} &
\colhead{$M_{\mathrm S}$} &
\colhead{$M_{\mathrm Ar}$} &
\colhead{$M_{\mathrm Ca}$} &
\colhead{$M_{\mathrm Fe}$} \\ 
\colhead{} &
\colhead{} &
\colhead{$({\mathrm M}_{\sun})$} &
\colhead{($10^{51}$\,erg)} &
\colhead{(mag)} &
\colhead{(mag)} &
\colhead{$({\mathrm M}_{\sun})$} &
\colhead{$({\mathrm M}_{\sun})$} &
\colhead{$({\mathrm M}_{\sun})$} &
\colhead{$({\mathrm M}_{\sun})$} &
\colhead{$({\mathrm M}_{\sun})$} &
\colhead{$({\mathrm M}_{\sun})$} 
}
\startdata
DET&
\nodata&
1.38&
1.59&
-19.87&
1.01&
0.0003&
0.0008&
0.0007&
0.0002&
0.0002&
1.22\\
SCH&
\nodata&
0.97&
1.01&
-17.53&
1.03&
0.09&
0.14&
0.10&
0.024&
0.026&
0.50\\
DEFa&
0.06&
1.37&
0.51&
-18.97&
0.85&
0.69&
0.025&
0.017&
0.0038&
0.0040&
0.55\\
DEFf&
0.16&
1.37&
0.84&
-19.43&
1.04&
0.48&
0.017&
0.012&
0.0028&
0.0029&
0.75\\
DDTa&
$3.9\times10^{7}$&
1.37&
1.40&
-19.73&
1.11&
0.04&
0.087&
0.071&
0.019&
0.022&
1.03\\
DDTe&
$1.3\times10^{7}$&
1.37&
1.02&
-19.00&
0.94&
0.19&
0.25&
0.19&
0.046&
0.054&
0.56\\
PDDa&
$4.4\times10^{7}$&
1.37&
1.45&
-19.79&
1.10&
0.02&
0.055&
0.045&
0.012&
0.015&
1.11\\
PDDe&
$7.7\times10^{6}$&
1.37&
1.12&
-19.02&
0.95&
0.10&
0.27&
0.22&
0.057&
0.067&
0.58\\
\enddata
\tablenotetext{a}{The parameter given is $\kappa$ for DEF models and
$\rho_{\mathrm tr}$ for DDT and PDD models (see the appendix for definitions)}    
\end{deluxetable}

\begin{deluxetable}{lcccc}
\tabletypesize{\scriptsize}
\tablecaption{Model spectral features at the age of Tycho SNR
\label{tab_linestrengths}}
\tablewidth{0pt}
\tablehead{
\colhead{Model} &
\colhead{Si K$\alpha $ line flux } &
\colhead{Fe K$\alpha $ line centroid\tablenotemark{a}} &
\colhead{$kT_{\mathrm sh}$\tablenotemark{b}} &
\colhead{$\left(EM_{\mathrm ISM}/4\pi D^{2}\right)$ \tablenotemark{b}} \\
\colhead{} &
\colhead{($10^{-3}$ph\,cm$^{-2}$\,s$^{-1}$)} &
\colhead{(keV)} &
\colhead{(keV)} &
\colhead{(cm$^{-5}$)} 
}
\startdata
HG97\tablenotemark{c} &
52.7&
6.458&
\nodata &
\nodata \\
DDTa&
11.95&
6.436&
27.4&
$1.39\times 10^{13}$\\
DDTe&
22.60&
----&
24.0&
$1.16\times 10^{13}$\\
PDDa&
23.22&
6.537&
28.4&
$1.45\times 10^{13}$\\
PDDe&
90.50&
6.465&
26.8&
$1.22\times 10^{13}$\\
DEFa&
3.46&
----&
22.6&
$0.58\times 10^{13}$\\
DEFf&
21.02&
----&
25.7&
$0.93\times 10^{13}$\\
DET&
1.89&
6.438&
27.9&
$1.48\times 10^{13}$\\
SCH&
31.35&
6.510&
25.1&
$1.21\times 10^{13}$\\
\enddata
\tablenotetext{a}{Centroids marked
as ---- would be impossible to determine from the model spectra due
to the weakness of the line}
\tablenotetext{b}{Parameters for the Sedov model for the shocked ISM}
\tablenotetext{c}{Observational results of \citet{hg}}
\end{deluxetable}

\begin{deluxetable}{lccccccc}
\tabletypesize{\scriptsize}
\tablecaption{Diagnostic line ratios for the models compared with the results 
of HG 97\label{tab_lineratios}}
\tablewidth{0pt}
\tablehead{
\colhead{Line Ratio} &
\colhead{Si K($\beta +\gamma $)/} &
\colhead{Si Ly$\alpha $/} &
\colhead{S K$\beta $/} &
\colhead{S Ly$\alpha $/} &
\colhead{S K$\alpha $/} &
\colhead{Ca K$\alpha $/} &
\colhead{Fe K$\alpha $/} \\
\colhead{} &
\colhead{Si K$\alpha $} &
\colhead{Si K$\alpha $} &
\colhead{S K$\alpha $} &
\colhead{S K$\alpha $} &
\colhead{Si K$\alpha $} &
\colhead{Si K$\alpha $} &
\colhead{Si K$\alpha $} 
}
\startdata
HG97 &
0.129&
0.028&
0.065&
$<0.010$&
0.26&
0.010&
0.008\\
DDTa&
0.090&
0.013&
0.011&
0.005&
0.55&
0.045&
0.006\\
DDTe&
0.105&
0.012&
0.012&
0.003&
0.33&
0.014&
0.001\\
PDDa&
0.161&
0.143&
0.068&
0.098&
0.49&
0.086&
0.029\\
PDDe&
0.162&
0.215&
0.088&
0.079&
0.50&
0.042&
0.001\\
DEFa&
----&
----&
----&
----&
0.34&
----&
----\\
DEFf&
0.147&
0.146&
----&
----&
0.27&
0.028&
----\\
DET&
0.069&
0.026&
----&
----&
0.49&
----&
0.074\\
SCH&
0.133&
0.175&
0.070&
0.050&
0.29&
0.007&
0.005\\
\enddata
\tablecomments{Ratios marked as ---- would be
impossible to determine from the model spectra due to the weakness
of the lines involved}
\end{deluxetable}

\begin{figure}
\epsscale{0.7}
\caption{\label{fig_composition}{\small Chemical composition profile of the
ejecta for the models DDTa (a), DDTe (b), PDDa (c), PDDe (d), DEFa
(e), DEFf (f), DET (g), and SCH (h). Elements are represented as follows:
C is the thin dash-triple-dotted plot (green in the electronic edition),
O is thick dash-triple-dotted (blue), Si is thick short-dashed (cyan),
S is thin short-dashed (yellow), Ar is long-dashed (purple), Ca is
dash-single-dotted (gray), Fe is thick solid (red) and Ni is dotted
(brown). The thin solid plot (black dotted plot in the electronic
edition) is the normalized density, 
\protect$\rho /\rho _{\mathrm max}\protect$.}}
\end{figure}

\begin{figure}
\caption{\label{fig_densitystart}{\small Density profiles for the supernova
models \protect$10^{7}\protect$\,s after the explosion. Model
DDTa is the thin solid line, DDTe
is dotted, PDDa is thin short-dashed, PDDe is dash-single-dotted, 
DEFa is dash-triple-dotted, DEFf is long-dashed,
DET is thick short-dashed, and SCH is thick solid.}}
\end{figure}

\begin{figure}
\caption{\label{fig_shockevol}{\small Temporal evolution of the shock parameters
\protect$r_{\mathrm fwd}\protect$ (a), \protect$r_{\mathrm rev}\protect$
(b), \protect$u_{\mathrm fwd}\protect$ (c), \protect$u_{\mathrm rev}\protect$
(d), \protect$\eta _{\mathrm fwd}\protect$ (e) and 
\protect$\eta _{\mathrm rev}\protect$ (f) for an interaction with 
\protect$\rho _{\mathrm AM}=10^{-24}\,{\mathrm g}\,{\mathrm cm}^{-3}\protect$.
Models marked as in Figure \ref{fig_densitystart}.}}
\end{figure}

\begin{figure}
\epsscale{0.7}
\caption{\label{fig_emevol} {\small \protect$EM_{X}/4\pi D^{2}\protect$
evolution for the models DDTa (a), DDTe (b), PDDa (c), PDDe (d), DEFa
(e), DEFf (f), DET (g), and SCH (h), for a distance of $D$=10\,kpc. The
thin solid plot (black in the electronic edition) corresponds to the
total \protect$EM/4\pi D^{2}\protect$ of the ejecta (i.e., their
total brightness). Elements labeled as in Figure \ref{fig_composition}.
C and O plots overlap in the deflagration models.}}
\end{figure}

\begin{figure}
\epsscale{0.7}
\caption{\label{fig_teemevol}{\small Emission measure averaged \protect$
T_{\mathrm e}\protect$
(\protect$\left\langle T_{\mathrm e}\right\rangle _{X}\protect$) for
the models DDTa (a), DDTe (b), PDDa (c), PDDe (d), DEFa (e), DEFf
(f), DET (g), and SCH (h). Elements labeled is as in Figure 
\ref{fig_composition}.
C and O overlap in the deflagration models.}}
\end{figure}

\begin{figure}
\epsscale{0.7}
\caption{\label{fig_tauemevol}{\small Emission measure averaged \protect$
\tau \protect$
(\protect$\left\langle \tau \right\rangle _{X}\protect$)for the
models DDTa (a), DDTe (b), PDDa (c), PDDe (d), DEFa (e), DEFf (f),
DET (g), and SCH (h). Elements labeled as in Figure \ref{fig_composition}.
C and O overlap in the deflagration models.}}
\end{figure}

\begin{figure}
\epsscale{0.7}
\caption{\label{fig_spectra}{\small Spatially integrated synthetic ejecta
spectra for the models DDTa (a), DDTe (b), PDDa (c), PDDe (d), DEFa
(e), DEFf (f), DET (g), and SCH (h). Spectra plotted with a solid
line correspond to 500 years after the explosion (blue in the electronic
edition), dotted line to 1000 years (red), dashed line to 2000 years
(orange) and dash-dotted line to 5000 years 
(green). Fluxes are for objects at a distance of $D$=10\,kpc. Spectra
are convolved with the response of the EPIC-MOS1 camera onboard the
XMM-Newton satellite. The \protect$K\alpha \protect$ lines of
Fe, Ca, S, and Si, as well as the O \protect$Ly\alpha \protect$
line, have been marked for clarity.}}
\end{figure}

\begin{figure}
\caption{\label{fig_Tychospectra}{\small Top: Spectrum of Tycho observed
by the XMM EPIC-MOS1 camera \citep[reproduced with permission]{d01}. 
Bottom: Predicted spectra for the models DDTa, PDDa,
PDDe and SCH. The solid line is the
total spectrum, the dotted line is the ejecta spectrum, and the
dashed line is the ISM spectrum.}}
\end{figure}


\begin{thebibliography}{}
\bibitem[Arnaud(1996)]{arn} Arnaud, K. A. 1996, in ASP Conf. Ser. 101, 
Astronomical
Data Analysis and Systems V, ed. G. Jacoby \& J. Barnes (San Francisco:
ASP), 17
\bibitem[Badenes and Bravo(2001)]{bb01} Badenes, C., \& Bravo, E. 2001, \apj, 
556, L41 
\bibitem[Badenes and Bravo(2003a)]{bb03a} Badenes, C., \& Bravo, E. 2003a, 
in ESA-SP 488, New Visions of the X-ray Universe in the XMM-Newton and Chandra 
Era, ed. F. Jansen (Noordwijk: ESA), in press (astro-ph/0202070)
\bibitem[Badenes and Bravo(2003b)]{bb03b} Badenes, C., \& Bravo, E. 2003b, 
in ESO Astrophysics Symposia, From Twilight
to Highlight: The Physics of Supernovae, eds. W. Hillebrandt \& B.
Leibundgut (Berlin: Springer), pp 264-267 (astro-ph/0209237)
\bibitem[Borkowski, Lyerly, and Reynolds(2001)]{blr} Borkowski, K. J., Lyerly, 
W. J., \& Reynolds, S. P. 2001, \apj, 548, 820
\bibitem[Branch and Khokhlov(1995)]{bk} Branch, D., \& Khokhlov, A. M. 1995, 
\physrep, 256, 53
\bibitem[Bravo et al.(1993)]{b93} Bravo, E., Dom\'\i nguez, I., Isern, J., 
Canal, R., Hofflich, P., \& Labay, J. 1993, \aap, 269, 187
\bibitem[Bravo and Garc\'\i a-Senz(1999)]{bg} Bravo, E., \& Garc\'\i a-Senz, D. 
1999, \mnras, 307, 984 
\bibitem[Bravo and Garc\'\i a-Senz(2003)]{bg3} Bravo, E., \& Garc\'\i a-Senz, 
D. 2003, in ESO Astrophysics Symposia, From Twilight
to Highlight: The Physics of Supernovae, eds. W. Hillebrandt \& B.
Leibundgut (Berlin: Springer), pp 165-168 (astro-ph/0211277)
\bibitem[Bravo et al.(1996)]{b96} Bravo, E., Tornamb\'e, A., Dom\'\i nguez, I., 
\& Isern, J. 1996, \aap, 306, 811 
\bibitem[Brinkmann et al.(1989)]{b89} Brinkmann, W., Fink, H. H., Smith, A., \& 
Habert, F. 1989, \aap, 221, 385
\bibitem[Cargill and Papadopoulos(1988)]{cp} Cargill, P. J., \& Papadopoulos, K.
 1988, \apj 329, L29
\bibitem[Decourchelle et al.(2001)]{d01} Decourchelle, A., et al. 2001, \aap, 
365, L218.
\bibitem[Dwarkadas and Chevalier(1998)]{dc} Dwarkadas, V. V., \& Chevalier, R. 
A. 1998, \apj, 497, 807 
\bibitem[Gamezo et al.(2003)]{g03} Gamezo, V.N., Khokhlov, A.M., Oran, E.S., 
Ctchelkanova, A.Y., \& Rosenberg, R.O. 2003, Science, 299, 77
\bibitem[Garc\'\i a-Senz and Bravo(2003)]{gb3} Garc\'\i a-Senz, D., \& Bravo, E.
2003, in ESO Astrophysics Symposia, From Twilight
to Highlight: The Physics of Supernovae, eds. W. Hillebrandt \& B.
Leibundgut (Berlin: Springer), pp 158-164 (astro-ph/0211242)
\bibitem[Ghavamian et al.(2001)]{g01} Ghavamian, P., Raymond, J., Smith, R. C., 
\& Hartigan, P. 2001, \apj, 547, 995
\bibitem[Gull(1973)]{gul} Gull, S. F. 1973, \mnras, 161, 47
\bibitem[Hamilton and Sarazin(1984)]{hs} Hamilton, A. J. S., \& Sarazin, C. L. 
1984, \apj, 287, 282
\bibitem[Hamilton, Sarazin, and Szymkowiak(1986)]{hss} Hamilton, A. J. S., 
Sarazin, C. L., \& Szymkowiak, A.E. 1986, \apj, 300, 698
\bibitem[Hendrick, Borkowski, and Reynolds(2003)]{hbr} Hendrick, S. P., 
Borkowski, K. J., \& Reynolds, S. P. 2003, \apj, submitted.
\bibitem[Hillebrandt and Niemeyer(2000)]{hn} Hillebrandt, W., \& Niemeyer, J. 
2000, \araa, 38, 191 
\bibitem[Hughes(2000)]{hug} Hughes, J. P. 2000, \apj, 545, L53 
\bibitem[Hughes et al.(2000)]{h00} Hughes, J. P., Rakowski, C. A., Burrows, D. 
N., \&  Slane, P. O. 2000, \apj, 528, L109
\bibitem[Hwang et al.(2002)]{h02} Hwang, U., Decourchelle, A., Holt, S.S. \& 
Petre, R. 2002, \apj, 581, 1101 
\bibitem[Hwang and Gotthelf(1997)]{hg} Hwang, U., \& Gotthelf, E. V. 1997, \apj,
475, 665
\bibitem[Hwang, Hughes, and Petre(1998)]{hhp} Hwang, U., Hughes, J. P., \& 
Petre, R. 1998, \apj, 497, 833
\bibitem[Itoh, Masai, and Nomoto(1988)]{imn} Itoh, H., Masai, K., \& Nomoto, K. 
1988, \apj, 334, 279
\bibitem[Khokhlov(2000)]{kho} Khokhlov, A.M. 2000, astro-ph/0008463
\bibitem[Laming(2000)]{l00} Laming, J. M. 2000, \apjs, 127, 409
\bibitem[Laming(2001)]{l01} Laming, J. M. 2001, \apj, 563, 828
\bibitem[Lewis et al.(2003)]{l03} Lewis, K. T., Burrows, D. N., Hughes, J. P., 
Slane, P. O., \& Garmire, G. P. 2003, \apj, 582, 770
\bibitem[Liedahl, Osterheld, and Goldestein(1995)]{log} Liedahl, D. A., 
Osterheld, A. L., \& Goldestein, W. H. 1995, \apj, 438, L115
\bibitem[Mazzotta et al.(1998)]{m98} Mazzotta, P., Mazzitelli, G., 
Colafrancesco, S., \& Vittorio, N. 1998, A\&AS, 133, 403
\bibitem[Niemeyer et al.(2003)]{n03} Niemeyer, J.C., Reinecke, M., Travaglio, 
C., \& Hillebrandt, W. 2003, in ESO Astrophysics Symposia, From Twilight
to Highlight: The Physics of Supernovae, eds. W. Hillebrandt \& B.
Leibundgut (Berlin: Springer), pp 151-157
\bibitem[Nomoto, Thielemann, and Yokoi(1984)]{nty} Nomoto, K., Thielemann, 
F.-K., \& Yokoi, K. 1984, \apj, 286, 644
\bibitem[Perlmutter et al.(1999)]{p99} Perlmutter, S., et al. 1999, \apj, 517, 
565
\bibitem[Schmidt et al.(1998)]{s98} Schmidt, B.P., et al. 1998, \apj, 507, 46
\bibitem[Summers and McWhirter(1978)]{sm} Summers, H.P., \& McWhirter, R.P. 
1978, J. Phys. B, 12, 2387
\bibitem[Thomas et al.(2002)]{t02} Thomas, R.C., Kasen, D., Branch, D., \&
Baron, E. 2002, \apj, 567, 1037
\bibitem[Timmes and Woosley(1992)]{tw} Timmes, F.S., \& Woosley, S.E. 1992, 
\apj, 396, 649
\bibitem[Truelove and McKee(1999)]{tm} Truelove, J.K., \& McKee, C.F. 1999, 
\apjs, 120, 299
\bibitem[van der Heyden et al.(2002)]{v02} van der Heyden, K. J., Behar, E., 
Vink, J., Rasmussen, A. P., Kaastra, J. S., Bleeker, J. A. M., Kahn, S. M. \& 
Mewe, R. 2002, \aap, 392, 955
\bibitem[Wang and Chevalier(2001)]{wc} Wang, C., \& Chevalier, R. 2001, \apj, 
549, 1119
\bibitem[Wheeler et al.(1987)]{w87} Wheeler, J.C., Swartz, D., Zong-wei, L., \&
Sutherland, P.G. 1987, \apj, 316, 733 
\bibitem[Wu et al.(1983)]{w83} Wu, C. C., Leventhal, M., Sarazin, C. L., \& 
Gull, T. R. 1983, \apj, 269, L5
\end{thebibliography}
\end{document}